\documentclass[prl,preprint,showpacs,preprintnumbers,amsmath,amssymb]{revtex4}

\usepackage[colorlinks=true,urlcolor=blue,citecolor=blue,linkcolor=blue]{hyperref}
\usepackage{graphicx}
\usepackage{dcolumn}
\usepackage{bm}
\usepackage{rotating}
\usepackage{lipsum}

\begin{document}

\title{Tunnel barrier design in donor nanostructures defined by hydrogen-resist lithography}

\author{Nikola Pascher$^1$, Szymon Hennel$^2$, Susanne Mueller$^2$, Andreas Fuhrer$^1$}

\affiliation{$^1$IBM Research - Zurich
S{\"a}umerstrasse 4,
8803 R{\"u}schlikon,
Switzerland \\
$^2$Solid State Physics Laboratory,
ETH Zurich,
8093 Zurich,
Switzerland}

\begin{abstract}
A four-terminal donor quantum dot (QD) is used to characterize potential barriers between degenerately doped nanoscale contacts. The QD is fabricated by hydrogen-resist lithography on Si(001) in combination with $n$-type doping by phosphine. The four contacts have different separations ($d$ = 9, 12, 16 and 29\;nm) to the central 6\;nm $\times$ 6\;nm QD island, leading to different tunnel and capacitive coupling. Cryogenic transport measurements in the Coulomb-blockade (CB) regime are used to characterize these tunnel barriers. We find that field enhancement near the apex of narrow dopant leads is an important effect that influences both barrier breakdown and the magnitude of the tunnel current in the CB transport regime. From CB-spectroscopy measurements, we extract the mutual capacitances between the QD and the four contacts, which scale inversely with the contact separation $d$. The capacitances are in excellent agreement with numerical values calculated from the pattern geometry in the hydrogen resist. We show that by engineering the source-drain tunnel barriers to be asymmetric, we obtain a much simpler excited-state spectrum of the QD, which can be directly linked to the orbital single-particle spectrum.  
\end{abstract}

\pacs{73.23.-b, 73.23.Hk, 73.63.Rt, 73.63.Kv, 68.37.Ef, 81.16.Nd, 68.47.Fg,}

\maketitle

\section{Introduction}
Using degenerately doped silicon is the most common way to contact the active region of many types of semiconductor devices. As device scaling requires ever smaller, more abrupt and more highly doped contacts, access resistance and dopant diffusion are critical issues in device performance and device variability. Hydrogen-resist lithography in combination with gas-phase doping provides a way to pattern degenerately doped contacts down to the nanometer scale with atomically abrupt dopant profiles \cite{94LydingAA,95ShenAB,02ShenAA,04RuessAA,05SimmonsAA}. This technique uses the tip of a scanning tunneling microscope (STM) as a lithography tool to locally remove a hydrogen passivation layer on the Si(001):H surface \cite{95ShenAB}. Exposure of those patterns to phosphine gas enables fabrication of extremely shallow and abrupt junctions with sheet densities on the order of $2\times10^{14}$\;cm$^{-2}$ \cite{02ShenAA,04RuessAA}. This enabled fabrication of phosphorus-doped wires with lithographic widths on the order of 1-2\;nm that exhibit Ohmic conduction at temperatures $<$ 100\;mK \cite{12WeberAA}. Furthermore, such wires were used to contact donor quantum dots containing a few or even a single dopant atom  \cite{09FuhrerAA,10FuechsleAA,12FuechsleAA}, with the ultimate goal of fabricating donor-based qubits.\\
A drawback of the hydrogen-resist technique is that the electronic coupling to the contacted object cannot be tuned much once the device has been fabricated \cite{09FuhrerAA,10FuechsleAA,14WeberAA}. Moreover, the reduced dimensionality of the contacts and the random arrangement of dopants within the patterned regions lead to a peaked density of states, which can affect cryogenic transport measurements such as bias spectroscopy in the Coulomb-blockade (CB) regime \cite{13RyuAA,10MottonenAA,10EscottAA}. It is thus important to understand how the geometry, separation and width of such degenerately doped contacts affect electrical transport.

\section{Characterization of the potential barriers}
We use a four-terminal quantum dot (QD) as an instrument to characterize potential barriers between highly doped nanoscale regions.  
\begin{figure}[]
\centering
	\includegraphics[width=0.7\columnwidth]{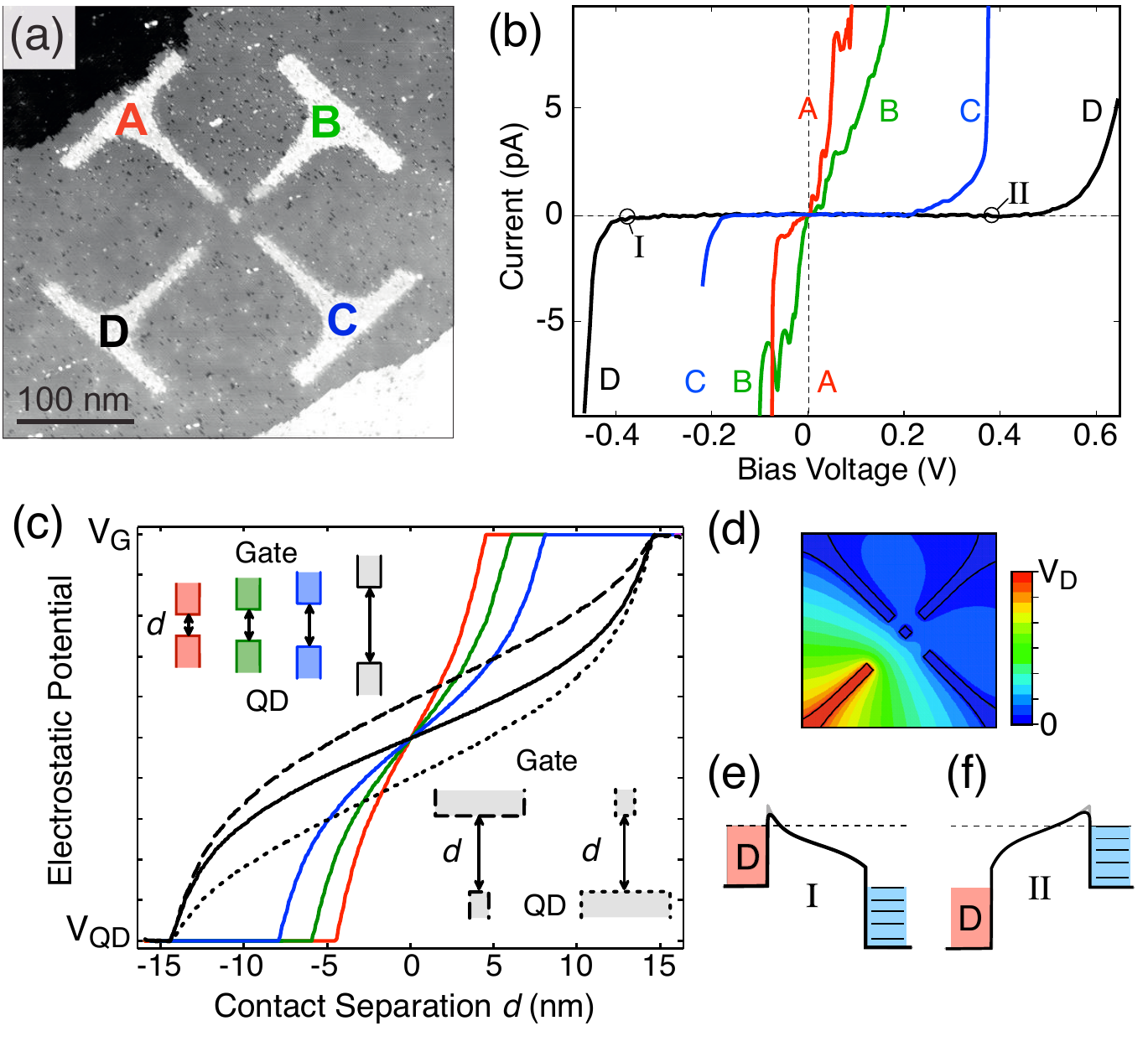}
	\caption{(a) Scanning tunneling micrograph of the QD structure during the lithography process. (b) Increase in barrier break-down voltage with larger contact separation. Current flow through each contact is measured as a function of the voltage applied to the same contact, while grounding all other connections. (c) Simulated electrostatic potential between two 6\;nm wide in-plane gate contacts with contact separations $d$ identical to thoase of the QD device. The four configurations are schematically shown in the top left inset using corresponding colors. For the largest separation (gray shading) the dashed/dotted black lines show the potential variation if the width of the upper/lower contact is increased to 30\;nm respectively. (d) Electrostatic potential in the QD plane when sweeping gate $D$. (e)+(f) Schematic potential drop between contact D and the QD for the two configurations $I$ and $II$ marked by circles in (b).}
	\label{Fig1}
\end{figure}
Figure 1 (a) shows an STM image of the device during the lithography process. The area where the hydrogen passivation layer has been removed appears bright in this image. During dosing of the sample with phosphine (6\;min at $5\times10^{-9}$\;mbar), the dopants only stick to the depassivated regions. After thermal incorporation of the dopants into the Si crystal (1\;min at 350$\,^\circ$C), the sample is overgrown with 20\;nm of undoped epitaxial Si. The silicon substrate is only lightly phosphorus-doped at $5\times10^{14}$\;cm$^{-3}$ and becomes insulating below a temperature of 40\;K. The sample is then removed from the UHV environment and Ohmic aluminum contacts are fabricated to the burried dopant device by electron-beam lithography and metal lift-off\cite{04RuessAA,09FuhrerAA}. \\
The donor QD is defined by a small (6\;nm $\times$ 6\;nm) island in the middle of the structure in Fig. 1 (a). The four electrodes A,B,C and D are patterned at distances of 9, 12, 16 and 29\;nm from the QD, respectively. These distances are designed to cover a range in which contacts A and B act as tunneling barriers, and contacts C and D act as gate electrodes. Measurements are performed in a dilution refrigerator at a base temperature of 50\;mK. A small magnetic field of 100\;mT is applied to inhibit superconductivity of the aluminum-bond pads. We use a source-measurement setup that enables us to set a voltage on each contact and simultaneously detect the currents in all four terminals. \\
Figure 1 (b) shows current-voltage traces for each of the contacts. Here, the bias voltage $V_X$ is varied and the corresponding current $I_X$ is plotted for all four contacts, with $X = A,B,C,D$. As expected, the onset of current leakage scales inversely with the patterned electrode distances $d$. For contacts A and B, current flows already at small bias voltages such that these contacts can be regarded as source and drain contacts of the QD. In contrast to this, leakage into contacts C and D only occurs for voltages $|V_X| > 0.2$\;V. Below this voltage, contacts C and D act as gates. \\
The breakdown of STM-defined potential barriers can be understood in a Fowler-Nordheim tunneling picture, where the applied bias leads to field-emission across a tilted potential barrier: 
\begin{equation}
I_X = c_1 E_X^2 e^{-c_2/|E_X|}= c_1 V_X^2/d^2 e^{-c_2 d/|V_X|}
\end{equation}
Here, $c_1$ and $c_2$ are constants that depend on the barrier height and charge-carrier mass. From this we expect a symmetric breakdown voltage that scales with $1/d$. However, the planar geometry and the pointed ends of our dopant contacts lead to deviations from the triangular barrier potential that is assumed for field emission. We numerically calculate the electrostatic potential between two 6\;nm wide planar contacts for the four different separations with ANSYS Maxwell, an electromagnetic field solver. The electrostatic potential drop in the low-doped region between the contact and the QD is shown in Fig.\;1\;(c). In addition to this, we expect a sharp step $\Phi_0$ in the potential at the interface between the highly doped contact and the intrinsic region in between. From the calculations in Ref.\; \cite{13RyuAA}, we estimate this barrier height $\Phi_0\approx E_C-E_F$ to be 80\;meV, where $E_C$ labels the energy of the conduction band edge and $E_F$ denotes the Fermi energy of the degenerately doped contact. For small bias ($V_{QD} \approx V_G$) the barrier is therefore roughly square-shaped. At negative gate voltages $V_G-V_{QD} < -\Phi_0$, the tunnel barrier potential becomes more triangular with a width determined by the electric field near the gate (see Fig.\;1\;(e)+(f)). Because of the planar geometry, the electric field lines are denser near the gate contact and the tunnel barrier is thinner than for the same situation in a conventional Fowler-Nordheim tunneling picture with a triangular barrier. \\
The observed asymmetry in barrier breakdown voltage (see, e.g., trace D in Fig.\;1\;(b)) was previously attributed to the depletion of thin dopant contacts under strong fields \cite{09FuhrerAA}. While this may play a role for extreme bias conditions, the asymmetry is more likely an effect of the point-like shape of the gate contacts. In Fig.\;1 (c), the dashed and dotted black lines are calculations for a situation where either gate D or the QD is made $5\times$ wider than the other contact (see lower right inset). In the patterned structure, the QD has the same lithographic width as contact D. However, as all other gates are grounded, the electrostatic potential distribution looks like the one shown in Fig.\;1\;(d), which is strongly asymmetric. Choosing $V_D=0.4$\;V it specifically also describes the situation at the two points $I$ and $II$ marked by circles in Fig.\;1\;(b). Figure\;1\;(e) shows the potential barrier for $I$ at negative gate bias. Here, the tunnel barrier is effectively thinner. Figure\;1\;(f) corresponds to situation $II$ for a positive gate bias and a thicker barrier. From our measurements, we find that the field at which contacts with a width of 6\;nm start showing barrier breakdown is about $15\pm4$\;mV/nm, with the sign depending on the polarity of the applied voltage. \\
We note that there are additional effects that will modify the shape of the barrier. The boundary of the doped regions will be blurred on the order of the effective Bohr radius, $a^*_B\approx3$\;nm. This will reduce the effective barrier width significantly for the smaller barriers. In close proximity to the doped regions, we may also expect a reduction of the potential due to image force effects. These will, however, be much smaller than for non-planar tunnel-barrier arrangements. 

\section{Current flow in the Coulomb-blockade regime}
Keeping in mind the properties of individual STM-defined potential barriers, we now probe current flow in the CB regime of the QD. For this, we fix $V_B=-10$\;mV and $V_C=10$\;mV, while sweeping $V_A$ as source drain bias and $V_D$ as gate voltage. Figure 2 shows the corresponding maps of the four simultaneously measured currents. Red denotes an electron current flow out of the contact and blue into the contact. We use a distorted color scale that highlights small currents; gray indicates regions in which the current exceeded the range of our measurement setup.\\
\begin{figure}[!tbp]
\centering
	\includegraphics[width=0.8\columnwidth]{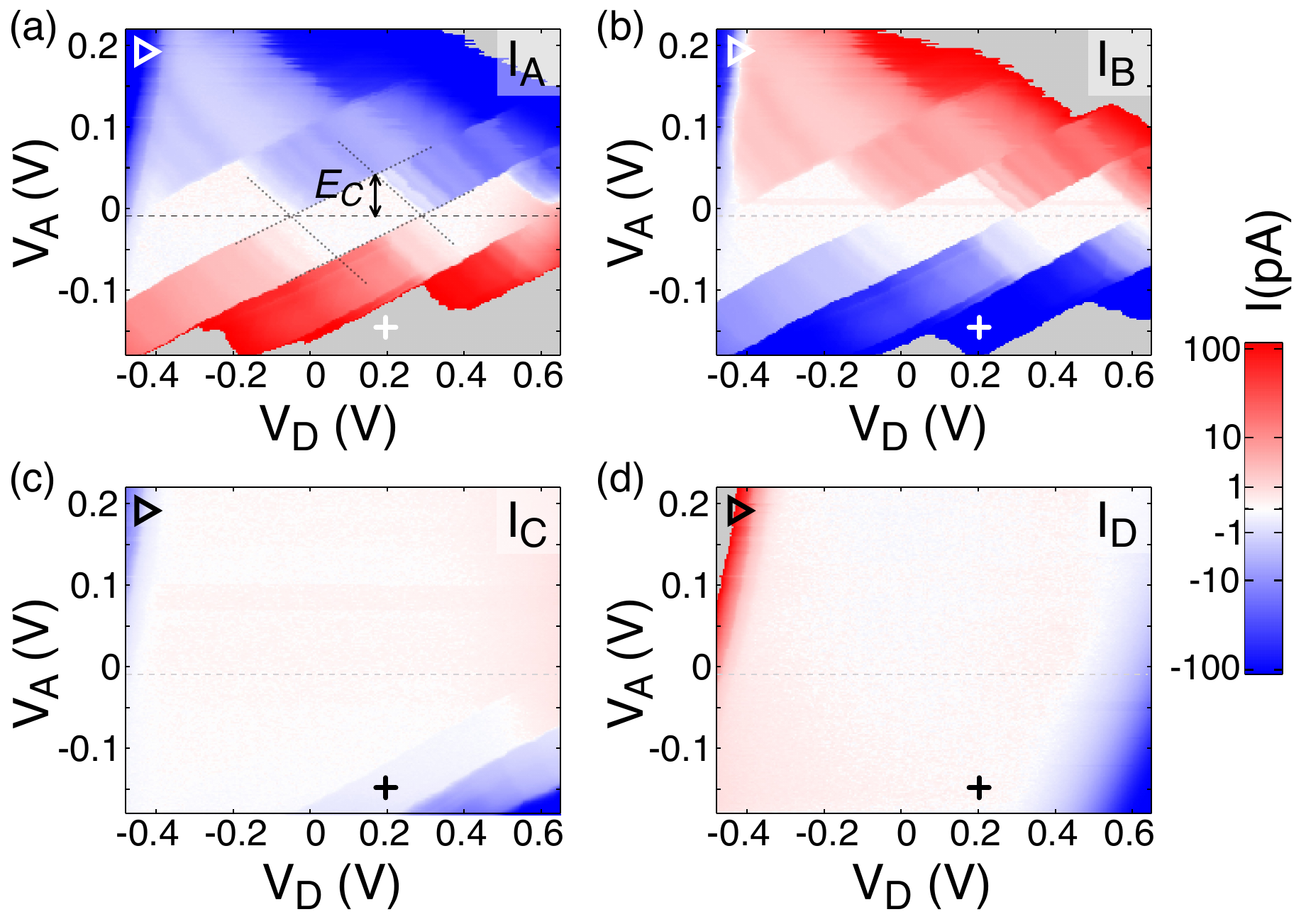}
	\caption{Current flowing in the four contacts as a function of $V_A$ and $V_D$. (a) and (b)  Conventional CB diamonds with $V_A$ as source-drain bias and $V_D$ as a gate voltage. (c) Leakage currents through contact C, which are modulated by the mean QD electron occupation (region marked with a cross). (d) Leakage current through contact D (e.g. region marked with a triangle). }
	\label{Fig2}
\end{figure}
In Figure 2\;(a) and (b), $I_A$ and $I_B$ show three CB diamonds consistent with CB transport through the QD from contact A to B and vice versa. The dashed zero bias line lies at $V_A=V_B=-10$\;mV and a charging energy $E_C = 50$\;meV can be determined from the extent of the diamonds in bias direction as indicated by the arrow in Fig. 2\;(a). The two boundary line slopes (dotted lines) defining the diamond edges are not symmetric. The sharper boundary, with a positive slope, corresponds to a situation in which the chemical potential of the more strongly coupled contact (here $\mu_A$) is fixed relative to the QD levels. Along lines that are parallel to this boundary, alignment of $\mu_B$ changes with respect to the QD levels. As contact B was designed with a larger gap, it limits the tunneling rate. The current modulation along such lines, beyond the CB region, originates either from transport through QD excited states or a modulated density of states in contact B \cite{09FuhrerAA,10FuechsleAA,10MottonenAA}.With an estimated 30 electrons trapped on the QD, we can change the electron number by about ten percent within the accessible voltage range in $V_D$. Leakage appears for $V_D < -0.4$\;V and $V_D > +0.45$\;V, as evidenced by $I_D$ in Fig. 2\;(d). For example, electron current leakage occurs from contact D into all other contacts near the area marked with a triangle in Fig.\;2. \\
An interesting situation occurs for leakage into contact C as shown in Fig. 2\;(c): In the voltage range near the cross symbol, electrons flow from contact A to both contacts B and C even though the coupling is expected to be significantly lower for contact C because of the larger separation. At the same time, $I_C$ is clearly modulated by the electron occupation of the QD determined by alignment with $\mu_A$, the electro chemical potential of contact A. As contact B is the rate-limiting contact, the average electron occupation of the QD increases whenever an additional electron is able to enter the QD from A. With increasing electron number the QD states that are being filled are a bit less localized and coupling to the contacts increases. This is seen clearest in the small leakage current into contact C. For the configuration marked with the cross symbol, electrons therefore tunnel sequentially from contact A into the QD and then into B or C with a probability ratio $\Gamma_B/\Gamma_C\approx I_B/I_C=10^{3}$. In the opposite current direction, for positive $V_A$, we observe no leakage in contact C. Here, the smaller electron occupation of the QD and the reduced effective gate voltage between D and the QD lead to a diminished $\Gamma_C$.

\begin{figure}[]
\centering
	\includegraphics[width=0.8\columnwidth]{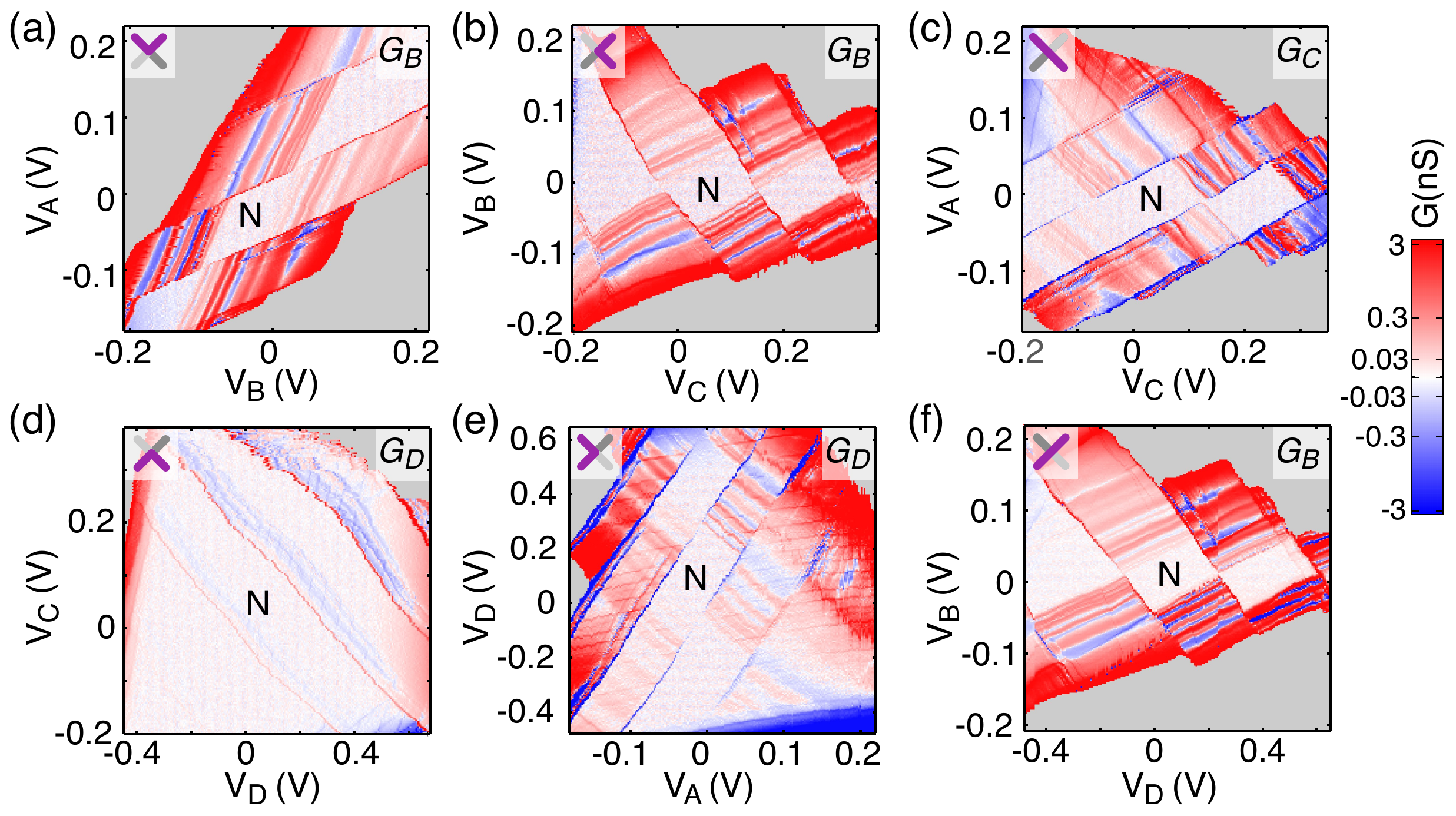}
	\caption{Differential conductance $G_X = \partial I_A/\partial V_X$ for all possible measurement configurations with two variable voltages. In (a),\;(b),\;(d) and (e), two neighboring gates are swept, indicated by the purple contacts in the top left corner of each plot. The two other voltages are fixed at +10\;mV (dark gray) and -10\;mV (light gray). For (c) and (f), the gates that are tuned oppose each other. }
	\label{Fig3}
\end{figure}

\section{Probing different contact configurations}
Instead of probing all four currents, we now focus our attention on current $I_A$. We choose contact A with the smallest gap to achieve the highest sensitivity to current flow through the QD. In Fig.\;3 the differential conductance $G_X = \partial I_A/\partial V_X$ for all measurement configurations with two variable voltages is shown. Fig.\;3\;(a),\;(b),\;(d) and (e) are maps as a function of two neighboring contact voltages. For Figs.\;3\;(c) and (f) the contacts with variable voltages oppose each other. The contacts that are tuned are highlighted in purple in the schematic at the top left corner of each plot. The remaining two contact voltages are again fixed at small values of $\pm10$\;mV. Figs.\;3\;(b),\;(c),\;(e) and (f) show clear CB-diamonds because both a gate and a bias voltage are swept. \\
In Fig.\;3\;(a) the QD is gated along one diagonal by the symmetric component of the two bias voltages $V_A+V_B$, whereas the source-drain bias across the dot is tuned by the asymmetric component $V_A-V_B$. Finally in Fig.\;3\;(d), the two gate voltages $V_C$ and $V_D$ are tuned in a gate-gate sweep. Here, the CB peaks show up as thin red lines and move diagonally from the top left to the bottom right corner. Both gate contacts are found to have a slight deficiency. At large positive $V_C$, the measurement is unstable even though there is no clear leakage into any of the other gates. Furthermore, whenever we tune $V_D$ additional faint lines appear with a fast period in $V_D$. We suspect that the latter is due to an area in contact D that was not well connected in the STM patterning step and now shows charging effects when $V_D$ is changed.\\
The two diamond plots (b) and (f) are both tuned by $V_B$ and look very similar except for a rescaled gate-voltage axis as a consequence of a smaller capacitive coupling of $V_D$ to the QD as compared to $V_C$. The same is true for the two diamond plots in (e) and (c). Here, the stronger capacitive coupling of $V_A$ to the QD also rescales the bias voltage axis in comparison to (b) and (f). We have indicated the Coulomb diamond with N electrons in all six panels of Fig.\;3. If the different lever arms of the four contacts are taken into account the four diamond plots look nearly identical, with the same excited state lines appearing. We therefore perform a more quantitative analysis of this using a constant interaction picture.\\ 

\begin{figure}[]
\centering
	\includegraphics[width=0.3\columnwidth]{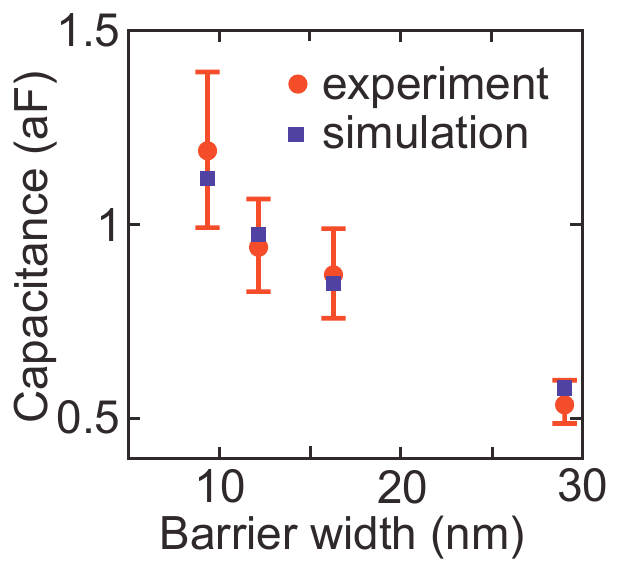}
	\caption{Comparison of mutual capacitances as a function of dot-contact separation. The red dots mark the experimental values determined from the CB diamond boundary slopes. Values denoted with the blue squares were calculated from the shape of the STM pattern assuming a metallic layer with a thickness of 1\;nm. }
	\label{Fig4}
\end{figure}

The chemical potential of the QD is given by
\begin{equation}
	\mu^Q_N = \epsilon^Q_N + \frac{e^2}{C_\Sigma}\left(N-\frac{1}{2} \right) + e \left( \frac{Q_{\infty}}{Q_{\Sigma}} + \sum_{j=1}^N \frac{C_{Q,j}}{C_{\Sigma}} V_j \right)
\label{energyisland}
\end{equation}
Here, $C_{\Sigma}=C_{QA}+C_{QB}+C_{QC}+C_{QD}$ is the sum over all mutual capacitances between gates and the QD. Using the experimental data, we find $C_{\Sigma} = e/E_C = 3.2$\;aF, which is similar to the value in other dopant QD devices \cite{06SellierAA,09PierreAA,10FuechsleAA,12FuechsleAA,13BuchAA}. Setting $\mu^Q_N=\mu^A=eV_A$ or $\mu^Q_N=\mu^B=eV_B$ allows us to express the diamond boundary line slopes $\frac{\partial V_X}{\partial V_Y}$ using the capacitances above and, by comparison with the experimentally observed slopes, to extract the mutual capacitances. These are shown as red dots in Fig.\;4\;(a) as a function of the separation $d$ between contacts and the QD. The error bars quantify the variation of the slopes in the experimental data. For comparison, we use the pattern in the STM image in Fig.\;1\;(a) as an input to the electromagnetic field solver and calculate the capacitance matrix numerically. For this, we assume metallic electrodes with a thickness of 1\;nm. The result is shown as blue squares in Fig.\;4\;(a). We find excellent agreement between the capacitances extracted from the measured diamonds and those calculated using the geometry of the STM pattern. This indicates that the nanoscale geometry of the donor device remains intact throughout all $in-situ$ and $ex-situ$ processing steps. 
\begin{figure}[]
\centering
	\includegraphics[width=0.8\columnwidth]{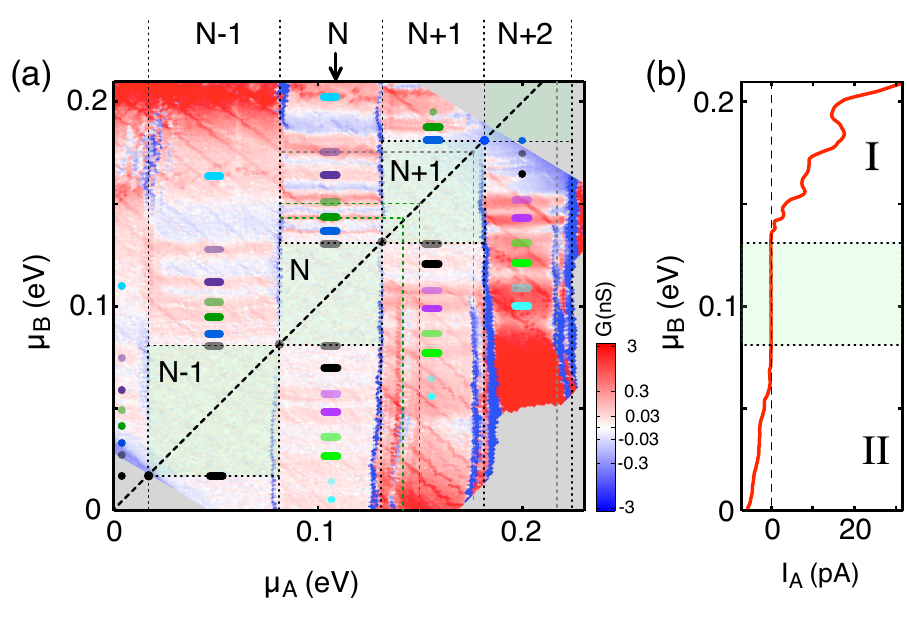}
	\caption{(a) QD excitation spectrum. Coulomb-blockaded regions are shaded in green. CB boundary lines are shown as thin dotted lines with the corresponding electron number given above the plot. Excited states are labeled by colored bars. (b) Vertical current trace for the N-electron state exhibiting negative differential conductance in the region denoted by I. }
	\label{Fig5}
\end{figure}

With the capacitive coupling of each of the contacts we can convert the applied voltages to chemical potential energies. This is shown in Fig.\;5\;(a). The color scale shows $G_D = \frac{\partial I_A}{\partial V_D}$ as in Fig.\;3\;(e), with the Coulomb-blockaded regions highlighted in green. We find three diamonds with a size close to $E_C$ and one that is about $10$\;meV larger. This difference is indicative of the level spacing in the QD with some uncertainty due to the fact that the dot capacitance $C_{\Sigma}$ decreases with decreasing electron occupation of the QD \cite{10FuechsleAA,09FuhrerAA}. The excited-state lines beyond the boundary of the CB region are due to alignment of QD states with $\mu_B$ as the latter is the current-limiting contact. The blue regions between the red excited state lines are regions of negative differential conductance (NDC). This is clarified by a vertical trace for N electrons on the QD (see arrow in Fig.\;5\;(a)). Figure\;5\;(b) shows the corresponding current trace in $I_A$. When the barrier potential is tilted by changing $\mu_B$ the transmission is tuned asymmetrically as discussed above. This is indicated by the roman numerals referring to the two situations shown in Figs.\;1\;(e) and (f) but now for contact B. In region I, the higher transmission of B renders the source-drain coupling of the QD more symmetric, allowing a larger current through the QD. The systematic appearance of NDC in this region is explained by a peaked density of states (DOS) in the 6\;nm wide source-drain contacts \cite{13RyuAA,10EscottAA,10FuechsleAA,10MottonenAA,09PierreAA}. For a contact with a flat DOS, we would expect a stepwise increase in the current except for cases in which a long-lived higher-lying excited state of the QD blocks transport \cite{93WeisAA}. In our case we see a reduction of the current by about 20\% immediately after each step, which is indicative of a peak in the DOS of contact B near $\mu_B$. This is also consistent with calculations for 6.1\;nm wide leads in Ref.\; \cite{13RyuAA} in which the DOS is found to vary on the same order of magnitude.\\
For the excited states in Fig.\;5\;(a), we first note that the asymmetric coupling of contacts A and B is expected to greatly simplify the observed excitation spectrum of the QD. The average occupation probability of the QD is tuned predominantly by $\mu_A$ because of its much stronger coupling to the QD. As the electron number N is thus practically constant along vertical traces, we expect to directly probe the single-particle energy spectrum of the QD, except where it is interrupted by the Coulomb gap. In other words, above the CB diamond with N electrons, we observe excitations of the N-electron QD state as usual. However, below the same diamond we expect to see hole-like excitations of the N-electron QD state rather than electron excitations of the (N-1)-electron state.\\
With this in mind we label the reappearing orbital levels in the single-particle spectrum of the QD with colored lines. We find clear evidence of paired levels below the CB diamonds with a splitting of about $10$\;meV and indicate this with two shades of the same color. For the electron excitations above the CB diamonds the pairing is not as clear and other pair assignments than the ones chosen may be possible. Places in which an orbital level is expected to reoccur without observing an excited state line are marked with a dot of the expected color.
The observed pairing is interesting in view of a possible valley-orbit splitting in donor-based Si QDs \cite{13BuchAA,10FuechsleAA,13ZwanenburgAA}. The magnitude of the splitting in our case is three times larger than the valley splitting previously observed for a donor cluster \cite{13BuchAA}.\\ 
We note that for the N+1 and N+2 electron state we also see vertical excited-state lines indicative of alignment with $\mu_A$ (highlighted by the dashed lines in Fig.\;5\;(a)). This could be explained by orbitals that are localized more closely to contact B and thus exhibit a more symmetric tunnel coupling to both leads.

\section{Conclusions}
We used an STM-defined four-terminal dopant QD with four different tunnel barriers to characterize the capacitive and the tunnel coupling of nanoscale dopant contacts. We showed that the field enhancement near the apex of narrow dopant leads is an important effect that influences both barrier breakdown and the magnitude of the tunnel current in the CB transport regime.\\
Using CB spectroscopy, we were able to determine the capacitive coupling of the contacts to the QD and found excellent agreement with capacitances simulated from the geometry patterned by STM. Furthermore, we showed that by engineering the source/drain tunnel barriers to be asymmetric, we obtain a much simpler excited-state spectrum of the QD, which can be directly linked to the orbital single-particle spectrum.\\
Our experiments demonstrate that although electrical tunability of STM-defined dopant devices may be more limited than that for top-gate-defined quantum structures, the atomic precision of STM patterning makes up for this. The STM allows us to engineer dopant devices at the sub-nanometer scale, which are shown to preserve their properties throughout the subsequent processing steps. 

\section{Acknowledgments}
The authors would like to thank Gian Salis, Daniel Widmer, Tomas Skeren, Andreas Kuhlmann and Klaus Ensslin for fruitful discussions and experimental support. Funding from the EU-FET grants SiSpin 323841, SiAM 610637, PAMS 610446 and from the Swiss NCCR QSIT is gratefully acknowledged.


\section*{References}

\begin{thebibliography}{10}
\expandafter\ifx\csname url\endcsname\relax
  \def\url#1{{\tt #1}}\fi
\expandafter\ifx\csname urlprefix\endcsname\relax\def\urlprefix{URL }\fi
\providecommand{\eprint}[2][]{\url{#2}}

\bibitem{94LydingAA}
Lyding J~W, Shen T~C, Hubacek J~S, Tucker J~R and Abeln G~C 1994  Nanoscale
  patterning and oxidation of h-passivated si(100)-2x1 surfaces with an
  ultrahigh vacuum scanning tunneling microscope {\em Appl. Phys. Lett.\/} {\bf
  64} 2010

\bibitem{95ShenAB}
Shen T~C, Wang C, Abeln G~C, Tucker J~R, Lyding J~W, Avouris P and Walkup E
  1995  Atomic-scale desorption through electronic and vibrational excitation
  mechanisms {\em Science\/} {\bf 268} 1590

\bibitem{02ShenAA}
Shen T~C, Ji J~Y, Zudow M~A, Du R~R, Kline J~S and Tucker J~R 2002  {Ultradense
  phosphorus $\delta$ layers grown into silicon from PH$_3$ molecular
  precursors} {\em Appl. Phys. Lett.\/} {\bf 80} 1580

\bibitem{04RuessAA}
Rue\ss F~J, Oberbeck L, Simmons M~Y, Goh K~E~J, Hamilton A~R, Hallam T,
  Schofield S~R, Curson N~J and Clark R~G 2004  Toward atomic-scale device
  fabrication in silicon using scanning probe microscopy {\em Nano Lett.\/}
  {\bf 4} 1969

\bibitem{05SimmonsAA}
Simmons M~Y, Ruess F~J, Goh K~E~J, Hallam T, Schofield S~R, Oberbeck L, Curson
  N~J, Hamilton A~R, Butscher M~J, Clark R~G and Reusch T~C~G 2005  Scanning
  probe microscopy for silicon device fabrication {\em Molecular Simulation\/}
  {\bf 31} 505

\bibitem{12WeberAA}
Weber B, Mahapatra S, Ryu H, Lee S, Fuhrer A, Reusch T, Thompson D, Lee W,
  Klimeck G, Hollenberg L~C {\em et~al.\/} 2012  Ohm's law survives to the
  atomic scale {\em Science\/} {\bf 335} 64

\bibitem{09FuhrerAA}
Fuhrer A, F{\"u}chsle M, Reusch T~C~G, Weber B and Simmons M~Y 2009
  Atomic-scale, all epitaxial in-plane gated donor quantum dot in silicon {\em
  Nano Lett.\/} {\bf 9} 707

\bibitem{10FuechsleAA}
Fuechsle M, Mahapatra S, Zwanenburg F~A, Friesen M, Eriksson M~A and Simmons
  M~Y 2010  Spectroscopy of few-electron single-crystal silicon quantum dots
  {\em Nature Nanotechnol.\/} {\bf 5} 502

\bibitem{12FuechsleAA}
Fuechsle M, Miwa J~A, Mahapatra S, Ryu H, Lee S, Warschkow O, Hollenberg L~C~L,
  Klimeck G and Simmons M~Y 2012  A single-atom transistor {\em Nature
  Nanotechnol.\/} {\bf 7} 242

\bibitem{14WeberAA}
Weber B, Ryu H, Tan Y~H~M, Klimeck G and Simmons M~Y 2014  Limits to metallic
  conduction in atomic-scale quasi-one-dimensional silicon wires {\em Phys.
  Rev. Lett.\/} {\bf 113} 246802

\bibitem{13RyuAA}
Ryu H, Lee S, Weber B, Mahapatra S, Hollenberg L~C~L, Simmons M~Y and Klimeck G
  2013  Atomistic modeling of metallic nanowires in silicon {\em Nanoscale\/}
  {\bf 5} 8666

\bibitem{10MottonenAA}
M{\"o}tt{\"o}nen M, Tan K~Y, Chan K~W, Zwanenburg F~A, Lim W~H, Escott C~C,
  Prikkalainen J~M, Morello A, Yang C, van Donkelaar J~A, Alves A~D~C, Jamieson
  D~N, Hollenberg L~C~L and Dzurak A~S 2010  Probe and control of the reservoir
  density of states in single-electron devices {\em Phy. Rev. B\/} {\bf 81}
  161304(R)

\bibitem{10EscottAA}
Escott C~C, Zwanenburg F~A and Morello A 2010  Resonant tunneling features in
  quantum dots {\em Nanotechnology\/} {\bf 21} 274018

\bibitem{06SellierAA}
Sellier H, Lansbergen G~P, Caro J, Rogge S, Collaert N, Ferain I, Jurczak M and
  Biesemans S 2006  Transport spectroscopy of a single dopant in a gated
  silicon nanowire {\em Phys Rev Lett\/} {\bf 97} 206805

\bibitem{09PierreAA}
Pierre M, Wacquez R, Jehl X, Sanquer M, Vinet M and Cueto O 2009  Single-donor
  ionization energies in a nanoscale cmos channel {\em Nature Nanotechnol.\/}
  {\bf 373} 133

\bibitem{13BuchAA}
B{\"u}ch H, Mahapatra S, Rahman R, Morello A and Simmons M~Y 2013  Spin readout
  and addressability of phosphorus-donor clusters in silicon {\em Nature
  Commun.\/} {\bf 4} 2017

\bibitem{93WeisAA}
Weis J, Haug R~J, Klitzing K~v and Ploog K 1993  Competing channels in
  single-electron tunneling through a quantum dot {\em Phys. Rev. Lett.\/} {\bf
  71} 4019

\bibitem{13ZwanenburgAA}
Zwanenburg F~A, Dzurak A~S, Morello A, Simmons M~Y, Hollenberg L~C~L, Klimeck
  G, Rogge S, Coppersmith S~N and Eriksson M~A 2013  Silicon quantum
  electronics {\em Rev. Mod. Phys.\/} {\bf 85} 961

\end{thebibliography}
\end{document}